\documentclass[prl,aps,amsmath,amssymb,showpacs,superscriptaddress,
twocolumn]{revtex4}
\usepackage{graphicx}
\usepackage{epsfig}
\usepackage{epsf}

\def\deltaf{\Delta_{\rm{eff}}}
\def\io{i\omega}
\def\Im{\rm{Im}}
\def\ek{\epsilon_{\bf k}}
\begin{document}

\def\beq{\begin{equation}} 
\def\eeq{\end{equation}} 
\def\be{\begin{equation}} 
\def\ee{\end{equation}} 
 
\def\iomn{i\omega_n} 
\def\iom#1{i\omega_{#1}} 
\def\c#1#2#3{#1_{#2 #3}} 
\def\cdag#1#2#3{#1_{#2 #3}^{+}} 
\def\epsk{\epsilon_{{\bf k}}} 
\def\Ga{\Gamma_{\alpha}} 
\def\Seff{S_{eff}} 
\def\dinf{$d\rightarrow\infty\,$} 
\def\T{\mbox{Tr}} 
\def\t{\mbox{tr}} 
\def\cG0{{\cal G}_0} 
\def\cS{{\cal S}} 
\def\divnum{\frac{1}{N_s}} 
\def\vac{|\mbox{vac}\rangle} 

\def\intR{\int_{-\infty}^{+\infty}} 
\def\intb{\int_{0}^{\beta}} 

\def\spinup{\uparrow} 
\def\spindown{\downarrow} 
\def\bra{\langle} 
\def\ket{\rangle} 
 
\def\ka{{\bf k}} 
\def\vk{{\bf k}} 
\def\vq{{\bf q}} 
\def\vQ{{\bf Q}} 
\def\vr{{\bf r}} 
\def\q{{\bf q}} 
\def\R{{\bf R}} 
\def\kp{\bbox{k'}} 

\def\a{\alpha} 
\def\b{\beta} 
\def\d{\delta} 
\def\D{\Delta} 
\def\e{\varepsilon} 
\def\ed{\epsilon_d} 
\def\ef{\epsilon_f} 
\def\g{\gamma} 
\def\G{\Gamma} 
\def\l{\lambda} 
\def\L{\Lambda} 
\def\o{\omega} 
\def\ph{\varphi} 
\def\s{\sigma} 

\def\chib{\overline{\chi}} 
\def\et{\widetilde{\epsilon}} 
\def\hn{\hat{n}} 
\def\hnu{\hat{n}_\uparrow} 
\def\hnd{\hat{n}_\downarrow} 
 
\def\hc{\mbox{h.c}} 
\def\Im{\mbox{Im}} 
 
\def\est{\varepsilon_F^*} 
\def\v2o3{V$_2$O$_3$} 
\def\uc2{$U_{c2}$} 
\def\uc1{$U_{c1}$}

\def\bea{\begin{eqnarray}} 
\def\eea{\end{eqnarray}} 
\def\#{\!\!}
\def\@{\!\!\!\!}
 
\def\vi{{\bf i}} 
\def\vj{{\bf j}} 

\def\+{\dagger}

\title{Is the Mott transition relevant to f-electron metals ?}

\author{L.~de' Medici}
\affiliation{Centre de Physique Th{\'e}orique, {\'E}cole Polytechnique
91128 Palaiseau Cedex, France}
\affiliation{Laboratoire de Physique des Solides, CNRS-UMR 8502, UPS
B{\^a}timent 510, 91405 Orsay, France}
\author{A.~Georges}
\affiliation{Centre de Physique Th{\'e}orique, {\'E}cole Polytechnique
91128 Palaiseau Cedex, France}
\author{G.~Kotliar}
\affiliation{Department of Physics and Astronomy, Serin Physics Laboratory,
Rutgers University, Piscataway 08854, USA}
\author{S.~Biermann}
\affiliation{Centre de Physique Th{\'e}orique, {\'E}cole Polytechnique
91128 Palaiseau Cedex, France}

\begin{abstract}
We study how a finite hybridization between a narrow correlated band and a wide
conduction band affects the Mott transition. At zero temperature, the hybridization is
found to be a relevant perturbation, so that the Mott transition is suppressed by Kondo
screening. In contrast, a first-order transition remains at finite temperature,
separating a local moment phase and a Kondo- screened phase. The first-order transition
line terminates in two critical endpoints. Implications for experiments on f-electron
materials such as the Cerium alloy Ce$_{0.8}$La$_{0.1}$Th$_{0.1}$ are discussed.
\end{abstract}

\maketitle

The Mott transition, i.e how electrons evolve from localized to itinerant
as a function of an external parameter such as pressure,
is a fundamental problem
in condensed matter physics. It is a key phenomenon in
$d$-electron materials, such as transition-metal oxides,
in which a set of bands with $d$-character is well separated
and close to the Fermi level.
In this case, the phase with localized electrons is insulating, and that
with itinerant electrons is metallic.
Dynamical mean-field (DMFT) studies~\cite{georges_review_dmft} have deepened  our
understanding of this phenomenon, and led to many interesting
experimental predictions which have been recently
verified experimentally~\cite{georges_mott_iscom}.

In several $f$-electron materials, a transition between a phase
where $f$-electrons are more localized and another in which they
are more itinerant is also observed (such as the isostructural
$\gamma$-$\alpha$ phase transition of
Cerium)~\cite{koskenmaki_gschneider,lawrence_repprogphys_1981}.
In these materials however, there is a broad band with $spd$-character close to the Fermi
level (in addition to the $f$-orbitals), and both phases are
metallic. It was suggested early on by
B.~Johansson~\cite{johansson_1974_mott_philmag} that the concept of a Mott
transition within the $f$-electron subspace may still be relevant
in this context. A different view is the Kondo volume collapse (KVC)
model~\cite{allen_martin_kvc_prl_1982,allen_liu_kvc_prb_1992,lavagna_kvc_physlett_1982},
in which the transition is driven by the change in the
hybridisation between the two phases with different unit-cell
volumes. In this picture, the broad band of conduction electrons
plays a key role, while it is merely a spectator in the Mott
picture.
The interplay of these two mechanims in Cerium have
recently received a great deal of attention, since DMFT provides a
comprehensive framework in which these problems can be studied
systematically~\cite{held_cerium_2001_prl,held_cerium_2003_prb,haule_cerium_prl_2005,Zolfl}.

In this article, we study the localization-delocalization transition within a
simple model, which nevertheless retains the key ingredients present in
$f$-electron materials. The model interpolates between a Hubbard model for the
$f$-orbital and the periodic Anderson model (PAM) in which this orbital is hybridized
to a broad band. Our goal is to understand whether, in a purely electronic model,
the Mott transition present when the $f$-band is isolated remains a robust feature
in the presence of a finite hybridization to a broad band.
Our key finding is that the answer to this question {\it depends on temperature}
in a crucial manner. At zero-temperature, the Kondo effect always sets in and
screens the local moment. As a result, in a purely electronic setting, the Mott transition
is {\it suppressed by an arbitrarily small hybridization}. In contrast, {\it a first-order
transition remains at finite temperature}.

The similarities
between the phase diagram of the PAM and that of the Hubbard model at finite
temperature have been pointed out in previous work~\cite{Held_PAM}. However, the
distinction between a first-order transition with coexisting electronic phases
and a mere crossover was not addressed. More importantly, the zero-temperature
case was investigated in the case where the hybridisation vanishes at the
Fermi level~\cite{Held_Bulla,vandongen_pam_prb_2001}: this is a non-generic
case in which the transition survives down to
$T=0$. In the generic case of a finite hybridisation, the connection between the
smooth behaviour at $T=0$ and the finite- temperature transition has not been
addressed before.
Our model study also has direct implications for the volume-collapse
transition of materials such as
Ce$_{0.8}$La$_{0.1}$Th$_{0.1}$~\cite{Thompson_Cerium_alloys},
and its dependence on magnetic field~\cite{Drymiotis_Ce_La_Th_Magnetic_field},
as explained at the end of this paper.

We study a generalisation of the periodic Anderson Model (PAM)
defined by the hamiltonian:
\bea\label{eq:hamiltonian}
H&=-t\sum_{\langle ij\rangle\s}c^\+_{i\s} c_{j\s} + V\sum_{i\s} c^\+_{i\s} f_{i\s}
-\a t\sum_{\langle ij \rangle\s}f^\+_{i\s} f_{j\s}  \nonumber \\
&\, + U \sum_i \left(n^f_{i\spinup} -\frac{1}{2}\right)
\left(n^f_{i\spindown}-\frac{1}{2}\right)
\eea
In addition to the usual hybridisation and interaction terms,
it contains a direct hopping between the $f$-orbitals: $t_{ff}=\alpha t$.
The model reduces to the PAM when $\alpha=0$. When $V=0$, it
describes two independent fluids: free conduction electrons,
and a narrow band of $f$-electrons described by the Hubbard model.
For simplicity, our study is restricted to the particle-hole symmetric
case where both bands are half-filled
($\langle n_{f}\rangle=\langle n_c\rangle =1$).
In this case, one has a (renormalised) hybridisation-gap insulator
when the direct $f$-$f$ hopping is small, as studied in \cite{shimizu_pamhop_jpsj_2000}
(for $\alpha=0$ and large $U$ this is the Kondo insulator).
For $\alpha$ large enough however, the hybridisation gap closes and the model describes a metal.
As shown below, the criterion for a metallic ground-state is essentially independent of
$U$ and reads: $\alpha>(V/D_c)^2$ (with $D_c$ the conduction electron bandwith).

In this article, we study this metallic regime
within DMFT, focusing on the paramagnetic phase.
When $V=0$, the situation is
well-documented~\cite{georges_review_dmft,georges_mott_iscom}: the f-electrons are described by
a Hubbard model which undergoes a Mott transition. The transition is first-order
at finite temperature, with a transition line $U_c(T)$ ending at a critical endpoint
$(U_c,T_c)$. The transition line separates two different regimes:
on one side the $f$-electrons are itinerant 
while on the other side the $f$-electrons
behave as local moments. These two behaviors correspond to two locally stable mean-field
solutions, which coexist in the domain $U_{c1}(T)<U<U_{c2}(T)$ delimited by
two spinodal lines.
The transition persists down to $T=0$: there, the quasiparticle weight
of the correlated itinerant solution vanishes continuously at $U_{c2}(T=0)$.
The main issue we want to address is whether a regime with unscreened local moments
survives in the presence of a finite hybridisation ($V\neq 0$)
to the broad conduction electron band, and what happens to the phase
transition.

DMFT associates to this lattice model a single-impurity Anderson model
for the $f$-orbital, subject to an effective hybridisation function
$\deltaf(\iomn)$ which must be determined self-consistently.
For the case of semi-circular densities of states for both the
$c$- and $f$-electrons (corresponding to a large-connectivity
Bethe lattice), this self-consistency condition can be written, using
the cavity construction~\cite{georges_review_dmft}, as:
\begin{equation}
\deltaf(\iomn)= \alpha^2t^2G_{ff}(\iomn) +
\frac{[V-\alpha t^2G_{cf}(\iomn)]^2}{\iomn-t^2G_{cc}(\iomn)}
\label{eq:delta_eff}
\end{equation}
In this expression, $G_{ff},G_{cf}$ and $G_{cc}$ are the different components of the
on-site interacting Green's function, which must be computed
self-consistently from the effective impurity model.
This expression has a transparent interpretation: The
screening of the $f$-moment on a given site in the local picture of the lattice
model has two origins reflected in each term of this equation. The first term
describes the screening due to the motion of the $f$-electrons onto other sites:
it is effective only when the $f$-electrons are itinerant, and its vanishing
at low-energy is associated with the Mott phenomenon. The second-term
describes the local screening due to the conduction electrons. This
screening is affected by the $f$-electron motion, resulting in a
reduced frequency-dependent effective hybridization
$V_{\rm{eff}}(\iomn)=V-\alpha t^2G_{cf}(\iomn)$.

\begin{figure}[h]
\includegraphics[width=7.5cm]{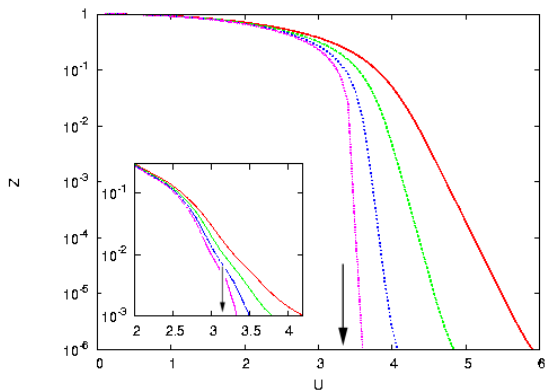}
\caption{\label{fig:cfr_Z_T=0} Comparison of Z vs U for different values of V
(from the left: 
$V=0.1, 0.2, 0.3, 0.4$) obtained using the
GA and $T=0$ ED (inset). Both methods clearly show that $Z$ never
vanishes as soon as $V\neq 0$, in contrast to the $V=0$ (Hubbard
model) case, which displays a transition at the critical value
of U indicated by the arrows.
Note that the GA overestimates~$Z$.}
\end{figure}

Let us consider first the case $T=0$.
Physical intuition suggests that an arbitrarily small hybridisation $V$
is enough to screen the local moment through the formation of a
Kondo singlet with the conduction electrons. The energy scale associated with
screening will be very small, but finite, at small $V$.
Hence, at $T=0$, the hybridization is a singular perturbation
when starting from the paramagnetic Mott phase, suggesting that
{\it the $T=0$ Mott transition is unstable} against the introduction
of hybridization.
This intuition is supported by a low-frequency analysis of
Eq.~(\ref{eq:delta_eff}). In a Mott phase with unquenched $f$-moments,
the Green's function and self-energy behave as:
$\Sigma_{f}(\io)\sim1/\io, G_{ff}(\io)\sim\io$ at small $\omega$.
Inserting this into (\ref{eq:delta_eff}), one sees that
$\deltaf(\io)\sim\io$ as $\o\rightarrow 0$ if $V=0$, which is consistent with the
original assumption of a local moment as it implies a gap in
the hybridisation density of states $\Im\deltaf(\omega+i0^+)$.
However, as soon as $V\neq 0$, $\deltaf(\io)$ tends to a
finite (imaginary) value as $\o\rightarrow 0$ because
of the second term in (\ref{eq:delta_eff}). This implies
a finite value of $\Im\deltaf(\omega+i0^+)$ at low-frequency, which
is inconsistent with a free local moment.
Hence, at $T=0$ and when $V\neq 0$, the self-energy has a local Fermi-liquid form
$\Sigma_{f}(\io)\sim \io (1-1/Z)+...$
for all values of $U$, with $Z$ the $f$-quasiparticle weight.
At large $U$, $Z$ is very small and sets the scale for screening.
This yields two quasiparticle bands, which read (neglecting lifetime effects):
$2\omega_\vk^\pm=(1+\alpha Z)\ek\pm [(1+\alpha Z)^2\ek^2+4ZV^2]^{1/2}$. This
corresponds to the non-interacting bandstructure, with renormalized
parameters: $\alpha_{\rm{eff}}=Z\alpha, V_{\rm{eff}}=\sqrt{Z}\,V$. It is
easily seen that a hybridisation gap is present only if
$V_{\rm{eff}}>\sqrt{\alpha_{\rm{eff}}}\,D_c$. The quasiparticle
weight $Z$ drops out from this criterion: hence, the two quasiparticle
bands overlap and one has a metal when $V<\sqrt{\alpha}\,D_c$,
independently of $U$ as announced above. Accordingly, it follows from this
low-frequency analysis that, at $T=0$,
the $f$-spectral function is pinned at $\omega=0$ to its non-interacting value:
$A_{ff}(0)=\alpha^{-1}\rho_0(V/\sqrt{\alpha})$, for all $U$, as long as
$0<V\leq\sqrt{\alpha}\,D_c$ (with $\rho_0$ the non-interacting density
of states of the conduction band). In this Fermi-liquid state, the ``large''
Fermi surface encounters $n_c+n_f=2$ electrons per site.
This low-frequency analysis can be illustrated by a simple calculation
using  the Gutzwiller approximation (GA). In this approach,
one optimizes a variational energy depending on the probability of
double occupancy $d$, and the quasiparticle residue is obtained\cite{kotliar_ruckenstein} as
$Z=16d(1/2-d)$. The results of this approximation for our model
are displayed in Fig.~\ref{fig:cfr_Z_T=0}.
This figure clearly shows that the Brinkman-Rice transition
(analogous to the $U_{c2}$ found in DMFT), at which $Z$ vanishes in
the Hubbard model ($V=0$), is no longer present at finite $V$. We showed
analytically that, because $V$ introduces a logarithmic singularity
($\propto Z\ln Z$) in the
variational energy at small $Z$, the minimum is always found at a finite value
of $Z$. This also allows us to estimate the behaviour of $Z$ at large
$U\gg U_{c2}$, which has the expected exponential suppression characteristic of the
Kondo effect: $Z\sim c_{\alpha,V} e^{-\pi U D_c/32 V^2}$ (the prefactor $c_{\alpha,V}$
depends only weakly on $\alpha$ and $V$).
We performed a full quantitative solution of the DMFT equations
at $T=0$, in order to confirm and extend this analysis.
For this purpose, we used an exact diagonalization (ED) scheme based
on the Lanczos algorithm and an adaptative discretization of the effective
bath degrees of freedom~\cite{georges_review_dmft}.
The Green's function and self-energy obtained from ED (not shown) do
obey $G(i\omega)\sim -i\pi\alpha^{-1}\rho_0(V/\sqrt{\alpha})\,,\,
\Sigma(i\omega)\sim i\omega(1-1/Z)$ at low-frequency for all $V\neq 0$,
from which we obtained the quasiparticle weight displayed in
Fig.~\ref{fig:cfr_Z_T=0}. Besides confirming the analysis above, we
also performed ED calculations for increasing and decreasing sweeps in $U$
in order to check that {\it no other solution} of the DMFT equation
is present at $T=0$ when $V\neq 0$,
besides the Fermi-liquid one with screened $f$-moments and a large Fermi surface.
This is in contrast
to the Hubbard model ($V=0$) which has a coexistence region $[U_{c1},U_{c2}]$
between a Mott-localized and an itinerant solution extending down to $T=0$.

\begin{figure}[h]
\includegraphics[width=7.5cm]{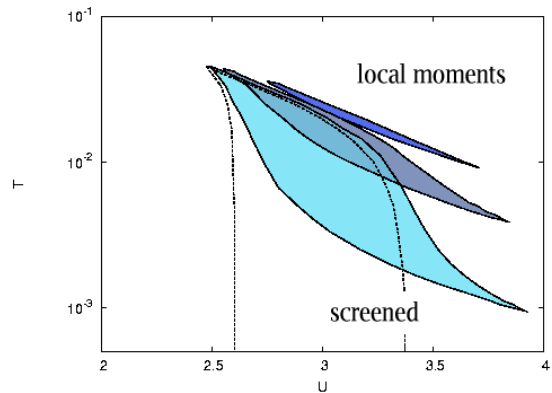}
\caption{\label{fig:PhaseDiagram_IPT} Phase diagram within the IPT approximation. Regions where localized and itinerant solutions coexist are 
displayed for (from bottom to top) V=0.1,0.2,0.3. 
Only finite-temperature transitions exist when $V\neq 0$,
resulting in two critical endpoints. In contrast, the spinodals
of the $V=0$ case (dashed lines) reach $T=0$ at finite critical values of $U$.}
\end{figure}

\begin{figure}[h]
\includegraphics[width=7.5cm]{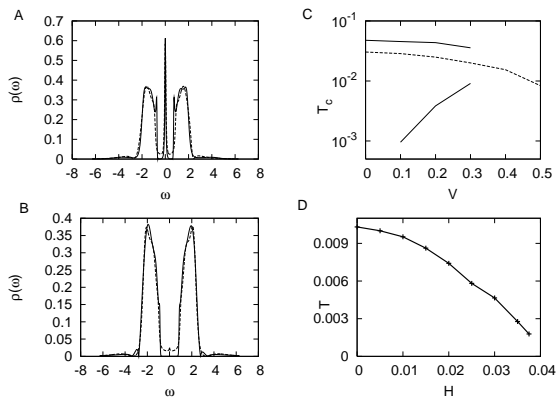}
\caption{\label{fig:rotors}
Left panel: comparison between spectral densities
for $V\neq 0$ (dashed lines) and $V=0$ (Hubbard Model, full lines),
in the screened phase~(A) and in the unscreened one~(B).
C:~Upper and lower $T_c$ of the critical endpoints as functions of
V within IPT
(full lines). Upper $T_c$ with DSR (dashed line).
D.~Transition temperature as a function of an
applied magnetic field H for U=2.7, calculated at $T=0$ within ED.
}
\end{figure}

While Kondo screening always sets in at $T=0$ and
suppresses the Mott transition, we expect the situation to
be qualitatively different at finite temperature. Indeed,
on general grounds, the effect of a perturbation (even if
singular for the ground-state) is expected to be smooth
at $T\neq 0$.
As a result, the first-order transition and the coexistence
region {\it should be robust features} of
the present model as long as $V$ is not too large.
Since small energy scales are involved ($T_c$ is of
order $\sim D_c/40$ for the pure Hubbard model, and the Kondo screening
scale is tiny at large $U$), an exact numerical study is
difficult and we approached the problem using two approximate
impurity solvers. The first is the iterated perturbation
theory (IPT) approximation\cite{georges_kotliar_dmft}
, which has proven to be semi-quantitatively
very successful in the study of the Mott transition. IPT is known to
overestimate low-energy scales, and will not be accurate in the Kondo
regime. The second method is
the (dynamical) ``slave-rotor'' (DSR) integral
equations\cite{florens_rotors_imp_2002_prb},
which is able to resolve low-energy scales and reproduces the
correct exponential Kondo scale at large $U$.
The phase diagram found within IPT is displayed in
Fig.~\ref{fig:PhaseDiagram_IPT}.
As anticipated, the coexistence (hysteretic) region is still present
for the smaller values of $V$.
As $V$ increases, its extension is drastically reduced, and $T_c$ decreases, as
also found with the
more accurate DSR solver (Fig.~\ref{fig:rotors} compares the estimates
of $T_c$ in the two methods). The spectral functions of two coexisting
solutions are displayed in Fig.~\ref{fig:rotors}: one has a
well-formed Kondo peak corresponding to good screening of the local moment while
the other one has very small (but finite) spectral weight at low energy.
In contrast to the $V=0$ case, we find that the two spinodals no longer
extend down to $T=0$ and that, within IPT, {\it another critical endpoint}
is found at low temperature at which the actual first-order transition line terminates.
Note that, in view of the above analysis at $T=0$, the two spinodal lines must indeed
either end at a lower critical point or run away towards infinite coupling.
Unfortunately, because of the low energy scale involved,
we have not been able to push the DSR method
to low enough temperatures and firmly establish the existence of a lower
critical endpoint within this technique.

We have also studied the effect of a magnetic field on the transition
between the screened (low-T, itinerant) and unscreened (high-T, local moment)
regimes.
This is motivated by recent experiments on the $\gamma-\alpha$ transition
of the Ce$_{0.8}$La$_{0.1}$Th$_{0.1}$ alloy, showing that the transition
temperature is decreased by an applied magnetic
field~\cite{Drymiotis_Ce_La_Th_Magnetic_field}.
Dzero {\it et al.}~\cite{Dzero} pointed out that this can be rationalized by approximating the
high-T $\gamma$- phase as a collection of almost free localized magnetic moments,
while assuming that the free-energy of the low-T $\alpha$-phase does not change
appreciably with magnetic field.
In Fig.~\ref{fig:rotors}D, we display our findings for the temperature associated with the
local moment spinodal line (lower boundary of the coexistence region), as
a function of applied field. This demonstrates, within the simple microscopic
model studied here, that indeed the transition is
suppressed by a magnetic field as observed experimentally.

To summarize, we have studied a model which retains the key aspects of f-electrons metals,
i.e. a narrow correlated band hybridized with a wide uncorrelated conduction band. We found a
first-order transition at finite temperature between a screened phase and a local-moment phase,
ending in two critical endpoints. We demonstrated that, at zero-temperature, the hybridization is
a relevant perturbation, so that the Mott transition is suppressed by Kondo screening. This is qualitatively consistent with the KVC picture.
These findings can be put in the broader context of the orbital-selective Mott transition (OSMT), which
attracted a lot of attention recently~\cite{Anisimov_OSMT,Koga_OSMT}.
In a general two-band case, the self-energy $\hat{\Sigma}$ takes a matrix form and the Mott transition is
signaled, when approached from the metallic side, by a low-frequency singularity in
$\omega.\hat{I}-\hat{\Sigma}(\omega)=\hat{Z}^{-1}\omega +\cdots$.
An OSMT is characterized by $\hat{Z}$ having one zero-eigenvalue while the other one
remains finite. We emphasize that this is a {\it basis-independent} notion. In our simple model,
$\hat{Z}$ is diagonal, with $Z_{cc}=1$, and hence an OSMT phase corresponds simply to the
vanishing of $Z_{ff}$. At $T=0$, this does occur for zero-hybridisation, but Kondo screening
prevents this to happen when a finite hybridization is turned on.

Finally, we comment on the qualitative relevance of our results
for f-electron materials. There, the contribution ($F_e$) to the
free-energy from the electronic degrees of freedom that are active
close to the transition have to be added to the contributions from
all other bands and ions, which can be approximated by an elastic
contribution (see e.g \cite{allen_martin_kvc_prl_1982}).
As a result, the volume-collapse transition does not
correspond to a true divergence of the response function
$\chi_e=-d^2F_e/dv^2$ of the active electronic degrees of freedom
(with $v$ the unit-cell volume).
Rather, it will take place\cite{hassan_sound_2004} when $\chi_e=B_0/v_0$, with $B_0$ and
$v_0$ a typical bulk modulus and unit-cell volume.
Hence the critical temperatures of the upper and lower
endpoints are shifted upwards and downwards, respectively, by
elastic terms. 
We therefore conclude that there are two generic
situations that are consistent with our results: either the
material displays a
first-order transition down to $T=0$ (for softer materials, with
smaller values of $B_0$) or it will display
two critical-endpoints (for
harder materials)(see also \cite{allen_liu_kvc_prb_1992}).
Experimental studies\cite{Thompson_Cerium_alloys}
suggest that the latter case may be realized in the Cerium alloy
Ce$_{0.8}$La$_{0.1}$Th$_{0.1}$, in which
alloying acts as a ``negative
pressure'', thus allowing for an investigation of the
localization-delocalization transition at lower temperatures than
in pure Cerium.
We emphasize that our results also imply that a $T=0$ quantum-critical
valence transition is a non-generic case that requires the
tuning of an extra parameter.

We thank J.~Allen, T.~Costi, M.~Rozenberg and M.~Vojta for useful discussions.
We acknowledge support from an RTN Network of the European
Union (HPRN-CT-2002-00295), from CNRS and from Ecole Polytechnique.
GK also acknowledges support by DOE Grant
No. LDRD-DR 200030084.


\begin{thebibliography}{25}
\expandafter\ifx\csname natexlab\endcsname\relax\def\natexlab#1{#1}\fi
\expandafter\ifx\csname bibnamefont\endcsname\relax
  \def\bibnamefont#1{#1}\fi
\expandafter\ifx\csname bibfnamefont\endcsname\relax
  \def\bibfnamefont#1{#1}\fi
\expandafter\ifx\csname citenamefont\endcsname\relax
  \def\citenamefont#1{#1}\fi
\expandafter\ifx\csname url\endcsname\relax
  \def\url#1{\texttt{#1}}\fi
\expandafter\ifx\csname urlprefix\endcsname\relax\def\urlprefix{URL }\fi
\providecommand{\bibinfo}[2]{#2}
\providecommand{\eprint}[2][]{\url{#2}}

\bibitem[{\citenamefont{{Georges} et~al.}(1996)\citenamefont{{Georges},
  {Kotliar}, {Krauth}, and {Rozenberg}}}]{georges_review_dmft}
\bibinfo{author}{\bibfnamefont{A.}~\bibnamefont{{Georges}}},
  \bibinfo{author}{\bibfnamefont{G.}~\bibnamefont{{Kotliar}}},
  \bibinfo{author}{\bibfnamefont{W.}~\bibnamefont{{Krauth}}}, \bibnamefont{and}
  \bibinfo{author}{\bibfnamefont{M.~J.} \bibnamefont{{Rozenberg}}},
  \bibinfo{journal}{Reviews of Modern Physics} \textbf{\bibinfo{volume}{68}},
  \bibinfo{pages}{13} (\bibinfo{year}{1996}).

\bibitem[{\citenamefont{{Georges} et~al.}(2004)\citenamefont{{Georges},
  {Florens}, and {Costi}}}]{georges_mott_iscom}
\bibinfo{author}{\bibfnamefont{A.}~\bibnamefont{{Georges}}},
  \bibinfo{author}{\bibfnamefont{S.}~\bibnamefont{{Florens}}},
  \bibnamefont{and} \bibinfo{author}{\bibfnamefont{T.~A.}
  \bibnamefont{{Costi}}}, \bibinfo{journal}{Journal de Physique IV -
  Proceedings} \textbf{\bibinfo{volume}{114}}, \bibinfo{pages}{165}
  (\bibinfo{year}{2004}), \eprint{cond-mat/0311520}.

\bibitem[{\citenamefont{Koskenmaki and
  Gschneider}(1978)}]{koskenmaki_gschneider}
\bibinfo{author}{\bibfnamefont{D.}~\bibnamefont{Koskenmaki}} \bibnamefont{and}
  \bibinfo{author}{\bibfnamefont{K.}~\bibnamefont{Gschneider}},
  \emph{\bibinfo{title}{Handbook on the Physics and Chemistry of Rare Earths}}
  (\bibinfo{publisher}{K.a. Gschneider and L. Eyring (North Holland,
  Amsterdam}, \bibinfo{year}{1978}).

\bibitem[{\citenamefont{J.~M.~Lawrence and
  Parks}(1981)}]{lawrence_repprogphys_1981}
\bibinfo{author}{\bibfnamefont{P.~M.~R.} \bibnamefont{J.~M.~Lawrence}}
  \bibnamefont{and} \bibinfo{author}{\bibfnamefont{R.~D.} \bibnamefont{Parks}},
  \bibinfo{journal}{Rep. Prog. Phys.} \textbf{\bibinfo{volume}{44}},
  \bibinfo{pages}{1} (\bibinfo{year}{1981}).

\bibitem[{\citenamefont{Johansson}(1974)}]{johansson_1974_mott_philmag}
\bibinfo{author}{\bibfnamefont{B.}~\bibnamefont{Johansson}},
  \bibinfo{journal}{Phil. Mag.} \textbf{\bibinfo{volume}{30}},
  \bibinfo{pages}{469} (\bibinfo{year}{1974}).

\bibitem[{\citenamefont{Allen and Martin}(1982)}]{allen_martin_kvc_prl_1982}
\bibinfo{author}{\bibfnamefont{J.~W.} \bibnamefont{Allen}} \bibnamefont{and}
  \bibinfo{author}{\bibfnamefont{R.~M.} \bibnamefont{Martin}},
  \bibinfo{journal}{Phys. Rev. Lett.} \textbf{\bibinfo{volume}{49}},
  \bibinfo{pages}{1106} (\bibinfo{year}{1982}).

\bibitem[{\citenamefont{Allen and Liu}(1992)}]{allen_liu_kvc_prb_1992}
\bibinfo{author}{\bibfnamefont{J.}~\bibnamefont{Allen}} \bibnamefont{and}
  \bibinfo{author}{\bibfnamefont{L.}~\bibnamefont{Liu}},
  \bibinfo{journal}{Phys. Rev. B} \textbf{\bibinfo{volume}{46}},
  \bibinfo{pages}{5047} (\bibinfo{year}{1992}).

\bibitem[{\citenamefont{Lavagna et~al.}(1982)\citenamefont{Lavagna, Lacroix,
  and Cyrot}}]{lavagna_kvc_physlett_1982}
\bibinfo{author}{\bibfnamefont{M.}~\bibnamefont{Lavagna}},
  \bibinfo{author}{\bibfnamefont{C.}~\bibnamefont{Lacroix}}, \bibnamefont{and}
  \bibinfo{author}{\bibfnamefont{M.}~\bibnamefont{Cyrot}},
  \bibinfo{journal}{Phys. Lett. A} \textbf{\bibinfo{volume}{90}},
  \bibinfo{pages}{210} (\bibinfo{year}{1982}).

\bibitem[{\citenamefont{{Held} et~al.}(2001)\citenamefont{{Held}, {McMahan},
  and {Scalettar}}}]{held_cerium_2001_prl}
\bibinfo{author}{\bibfnamefont{K.}~\bibnamefont{{Held}}},
  \bibinfo{author}{\bibfnamefont{A.~K.} \bibnamefont{{McMahan}}},
  \bibnamefont{and} \bibinfo{author}{\bibfnamefont{R.~T.}
  \bibnamefont{{Scalettar}}}, \bibinfo{journal}{Phys. Rev. Lett.}
  \textbf{\bibinfo{volume}{87}}, \bibinfo{pages}{276404}
  (\bibinfo{year}{2001}).

\bibitem[{\citenamefont{{McMahan} et~al.}(2003)\citenamefont{{McMahan}, {Held},
  and {Scalettar}}}]{held_cerium_2003_prb}
\bibinfo{author}{\bibfnamefont{A.~K.} \bibnamefont{{McMahan}}},
  \bibinfo{author}{\bibfnamefont{K.}~\bibnamefont{{Held}}}, \bibnamefont{and}
  \bibinfo{author}{\bibfnamefont{R.~T.} \bibnamefont{{Scalettar}}},
  \bibinfo{journal}{Phys. Rev. B} \textbf{\bibinfo{volume}{67}},
  \bibinfo{pages}{075108} (\bibinfo{year}{2003}).

\bibitem[{\citenamefont{Haule et~al.}(2005)\citenamefont{Haule, V.Oudovenko,
  Savrasov, and Kotliar}}]{haule_cerium_prl_2005}
\bibinfo{author}{\bibfnamefont{K.}~\bibnamefont{Haule}},
  \bibinfo{author}{\bibnamefont{V.Oudovenko}},
  \bibinfo{author}{\bibfnamefont{S.}~\bibnamefont{Savrasov}}, \bibnamefont{and}
  \bibinfo{author}{\bibfnamefont{G.}~\bibnamefont{Kotliar}},
  \bibinfo{journal}{Phys. Rev. Lett.} \textbf{\bibinfo{volume}{94}},
  \bibinfo{pages}{036401} (\bibinfo{year}{2005}).

\bibitem[{\citenamefont{Z\"olfl et~al.}(2001)\citenamefont{Z\"olfl, Nekrasov,
  Pruschke, Anisimov, and Keller}}]{Zolfl}
\bibinfo{author}{\bibfnamefont{M.~B.} \bibnamefont{Z\"olfl}},
  \bibinfo{author}{\bibfnamefont{I.~A.} \bibnamefont{Nekrasov}},
  \bibinfo{author}{\bibfnamefont{T.}~\bibnamefont{Pruschke}},
  \bibinfo{author}{\bibfnamefont{V.~I.} \bibnamefont{Anisimov}},
  \bibnamefont{and} \bibinfo{author}{\bibfnamefont{J.}~\bibnamefont{Keller}},
  \bibinfo{journal}{Phys. Rev. Lett.} \textbf{\bibinfo{volume}{87}},
  \bibinfo{pages}{276403} (\bibinfo{year}{2001}).

\bibitem[{\citenamefont{Held et~al.}(2000)\citenamefont{Held, Huscroft,
  Scalettar, and McMahan}}]{Held_PAM}
\bibinfo{author}{\bibfnamefont{K.}~\bibnamefont{Held}},
  \bibinfo{author}{\bibfnamefont{C.}~\bibnamefont{Huscroft}},
  \bibinfo{author}{\bibfnamefont{R.}~\bibnamefont{Scalettar}},
  \bibnamefont{and} \bibinfo{author}{\bibfnamefont{A.}~\bibnamefont{McMahan}},
  \bibinfo{journal}{Phys. Rev. Lett.} \textbf{\bibinfo{volume}{85}},
  \bibinfo{pages}{373} (\bibinfo{year}{2000}).

\bibitem[{\citenamefont{Held and Bulla}(2000)}]{Held_Bulla}
\bibinfo{author}{\bibfnamefont{K.}~\bibnamefont{Held}} \bibnamefont{and}
  \bibinfo{author}{\bibfnamefont{R.}~\bibnamefont{Bulla}},
  \bibinfo{journal}{Eur. Phys. J. B} \textbf{\bibinfo{volume}{17}},
  \bibinfo{pages}{7} (\bibinfo{year}{2000}).

\bibitem[{\citenamefont{{van Dongen} et~al.}(2001)\citenamefont{{van Dongen},
  {Majumdar}, {Huscroft}, and {Zhang}}}]{vandongen_pam_prb_2001}
\bibinfo{author}{\bibfnamefont{P.}~\bibnamefont{{van Dongen}}},
  \bibinfo{author}{\bibfnamefont{K.}~\bibnamefont{{Majumdar}}},
  \bibinfo{author}{\bibfnamefont{C.}~\bibnamefont{{Huscroft}}},
  \bibnamefont{and} \bibinfo{author}{\bibfnamefont{F.}~\bibnamefont{{Zhang}}},
  \bibinfo{journal}{Phys. Rev. B}
  \textbf{\bibinfo{volume}{64}}(\bibinfo{number}{19}), \bibinfo{pages}{195123}
  (\bibinfo{year}{2001}).

\bibitem[{\citenamefont{Thompson et~al.}(1983)\citenamefont{Thompson, Fisk,
  Lawrence, Smith, and Martin}}]{Thompson_Cerium_alloys}
\bibinfo{author}{\bibfnamefont{J.}~\bibnamefont{Thompson}},
  \bibinfo{author}{\bibfnamefont{Z.}~\bibnamefont{Fisk}},
  \bibinfo{author}{\bibfnamefont{J.}~\bibnamefont{Lawrence}},
  \bibinfo{author}{\bibfnamefont{J.}~\bibnamefont{Smith}}, \bibnamefont{and}
  \bibinfo{author}{\bibfnamefont{R.}~\bibnamefont{Martin}},
  \bibinfo{journal}{Phys. Rev. Lett.} \textbf{\bibinfo{volume}{50}},
  \bibinfo{pages}{1081} (\bibinfo{year}{1983}).

\bibitem[{\citenamefont{F.Drymiotis and al.
  condmat/0406043}(2004)}]{Drymiotis_Ce_La_Th_Magnetic_field}
\bibinfo{author}{\bibnamefont{F.Drymiotis}} \bibnamefont{and}
  \bibinfo{author}{\bibnamefont{al. condmat/0406043}}  (\bibinfo{year}{2004}).

\bibitem[{\citenamefont{Shimizu et~al.}(2000)\citenamefont{Shimizu, Sakai, and
  Hewson}}]{shimizu_pamhop_jpsj_2000}
\bibinfo{author}{\bibfnamefont{Y.}~\bibnamefont{Shimizu}},
  \bibinfo{author}{\bibfnamefont{O.}~\bibnamefont{Sakai}}, \bibnamefont{and}
  \bibinfo{author}{\bibfnamefont{A.~C.} \bibnamefont{Hewson}},
  \bibinfo{journal}{J. Phys. Soc. Jpn.} \textbf{\bibinfo{volume}{69}},
  \bibinfo{pages}{1777} (\bibinfo{year}{2000}).

\bibitem[{\citenamefont{Kotliar and Ruckenstein}(1986)}]{kotliar_ruckenstein}
\bibinfo{author}{\bibfnamefont{G.}~\bibnamefont{Kotliar}} \bibnamefont{and}
  \bibinfo{author}{\bibfnamefont{A.}~\bibnamefont{Ruckenstein}},
  \bibinfo{journal}{Phys. Rev. Lett.} \textbf{\bibinfo{volume}{57}},
  \bibinfo{pages}{1362} (\bibinfo{year}{1986}).

\bibitem[{\citenamefont{{Georges} and {Kotliar}}(1992)}]{georges_kotliar_dmft}
\bibinfo{author}{\bibfnamefont{A.}~\bibnamefont{{Georges}}} \bibnamefont{and}
  \bibinfo{author}{\bibfnamefont{G.}~\bibnamefont{{Kotliar}}},
  \bibinfo{journal}{Phys. Rev. B} \textbf{\bibinfo{volume}{45}},
  \bibinfo{pages}{6479} (\bibinfo{year}{1992}).

\bibitem[{\citenamefont{{Florens} and
  {Georges}}(2002)}]{florens_rotors_imp_2002_prb}
\bibinfo{author}{\bibfnamefont{S.}~\bibnamefont{{Florens}}} \bibnamefont{and}
  \bibinfo{author}{\bibfnamefont{A.}~\bibnamefont{{Georges}}},
  \bibinfo{journal}{Phys. Rev. B} \textbf{\bibinfo{volume}{66}},
  \bibinfo{pages}{165111} (\bibinfo{year}{2002}).

\bibitem[{\citenamefont{Dzero et~al.}(2000)\citenamefont{Dzero, Gor'kov, and
  Zvezdin}}]{Dzero}
\bibinfo{author}{\bibfnamefont{M.}~\bibnamefont{Dzero}},
  \bibinfo{author}{\bibfnamefont{L.}~\bibnamefont{Gor'kov}}, \bibnamefont{and}
  \bibinfo{author}{\bibfnamefont{A.}~\bibnamefont{Zvezdin}},
  \bibinfo{journal}{J. Phys.: Condens. Matter} \textbf{\bibinfo{volume}{12}}
  (\bibinfo{year}{2000}).

\bibitem[{\citenamefont{Anisimov et~al.}(2002)\citenamefont{Anisimov, Nekrasov,
  Kondakov, Rice, and Sigrist}}]{Anisimov_OSMT}
\bibinfo{author}{\bibfnamefont{V.}~\bibnamefont{Anisimov}},
  \bibinfo{author}{\bibfnamefont{I.}~\bibnamefont{Nekrasov}},
  \bibinfo{author}{\bibfnamefont{D.}~\bibnamefont{Kondakov}},
  \bibinfo{author}{\bibfnamefont{T.}~\bibnamefont{Rice}}, \bibnamefont{and}
  \bibinfo{author}{\bibfnamefont{M.}~\bibnamefont{Sigrist}},
  \bibinfo{journal}{Eur. Phys. J. B} \textbf{\bibinfo{volume}{25}},
  \bibinfo{pages}{191} (\bibinfo{year}{2002}).

\bibitem[{\citenamefont{Koga et~al.}(2004)\citenamefont{Koga, N.Kawakami, Rice,
  and Sigrist}}]{Koga_OSMT}
\bibinfo{author}{\bibfnamefont{A.}~\bibnamefont{Koga}},
  \bibinfo{author}{\bibnamefont{N.Kawakami}},
  \bibinfo{author}{\bibfnamefont{T.}~\bibnamefont{Rice}}, \bibnamefont{and}
  \bibinfo{author}{\bibfnamefont{M.}~\bibnamefont{Sigrist}},
  \bibinfo{journal}{Phys. Rev. Lett.} \textbf{\bibinfo{volume}{92}},
  \bibinfo{pages}{216402} (\bibinfo{year}{2004}).

\bibitem[{\citenamefont{{Hassan} et~al.}(2005)\citenamefont{{Hassan},
  {Georges}, and {Krishnamurthy}}}]{hassan_sound_2004}
\bibinfo{author}{\bibfnamefont{S.~R.} \bibnamefont{{Hassan}}},
  \bibinfo{author}{\bibfnamefont{A.}~\bibnamefont{{Georges}}},
  \bibnamefont{and} \bibinfo{author}{\bibfnamefont{H.~R.}
  \bibnamefont{{Krishnamurthy}}}, \bibinfo{journal}{Phys. Rev. Lett.}
  \textbf{\bibinfo{volume}{94}}, \bibinfo{pages}{036402}
  (\bibinfo{year}{2005}).

\end{thebibliography}
\end{document}